\documentclass[twocolumn,english,showpacs]{revtex4}
\usepackage{babel,amsmath,amssymb,dcolumn}
\usepackage[dvips]{graphics}

\begin{document}

\title{Massive vector field perturbations in the Schwarzschild
background: stability and quasinormal spectrum.}

\author{R.A. Konoplya}
\email{konoplya@fma.if.usp.br}
\affiliation{Instituto de F\'{\i}sica, Universidade de S\~{a}o Paulo \\
C.P. 66318, 05315-970, S\~{a}o Paulo-SP, Brazil}

\pacs{04.30.Nk,04.50.+h}

\begin{abstract}
We consider the perturbations of the massive vector field around
Schwarzschild, Schwarzschild - de Sitter, and  Schwarzschild - anti - de Sitter
black holes. Equations for a spherically symmetric massive vector 
perturbation can be reduced to a
single wave-like equation. We have proved the stability against these
perturbations and investigated the quasinormal spectrum. 
The quasinormal behaviour for Schwarzschild
black hole is quite unexpected: the fundamental mode 
and  higher overtones shows totally different dependence on the mass of
the field $m$: as $m$ is increasing, the damping rate 
of the fundamental mode is decreasing, what results in appearing 
of the infinitely long living modes, while, on contrary, damping rate of all higher
overtones are increasing, and their real oscillation frequencies gradually
go to tiny values. Thereby, for all higher overtones, 
almost non-oscillatory, damping modes can exist. In the limit 
of asymptotically high damping, $Re \omega$ goes to $ln3/(8 \pi M)$, while imaginary part shows
equidistant behaviour with spacing $Im \omega_{n+1}- Im
\omega_{n}=1/4M$. In addition, we have found quasinormal spectrum of massive vector
field for  Schwarzschild-anti-de Sitter  black hole. 
\end{abstract}

\maketitle

\section{Introduction}

The existence of the charged black hole described by the 
Reissner-Nordstrom solution reflects the fact that a black hole can
possess a massless monopole  vector (electromagnetic) field. It gets
rid of all higher multipoles through radiative processes dominated by
quasinormal ringing at intermediately late times and by power-low or
exponential tails at asymptotically late times.   

Since Bekenstein's paper \cite{1}, it is well-known that a black hole
cannot possess even a monopole massive vector field. Therefore 
the black hole has to radiate away the massive vector field with some 
quasinormal frequencies governing this radiation. Nevertheless, the massive
vector quasinormal modes of a  Schwarzschild black hole were not
studied so far, and, as we shall show in this paper, the problem
is qualitively different from that for a massive scalar field, leading to quite 
unusual quasinormal behaviour. First of all, let us briefly review 
what we know about massive scalar and massless vector field perturbations.  
    
The massless vector perturbations of the  Schwarzschild background was 
considered for the first time in \cite{2}. There  the 
Maxwell field perturbations were reduced to a single wave-like equation for some gauge
invariant function $\Psi =\Psi (r, t)$, 
\begin{equation}
\Psi_{, r_{*} r_{*}} - \Psi_{, t t} - 
\left(1-\frac{2 M}{r}\right) \frac{\ell (\ell+1)}{r^2} \Psi =0.   
\end{equation} 
Here $M$ is the black hole mass and $\ell$ is the multipole number.
Note, that this equation is valid only for $\ell>0$, while for $\ell=0$
(monopole, or spherically symmetrical perturbations) the Maxwell 
equations in Schwarzschild background do not exhibit dynamical degrees
of freedom. This signifies about existence of non-radiative
electromagnetic monopole hair, i.e. about existence of a black hole charge.
The quasinormal modes and late time behaviour stipulated by this effective potential were
found in a lot of papers (see recent papers 
\cite{3}, \cite{siopsis}, \cite{Konoplya:2004ip}, \cite{3a} and
references therein), and are well-studied.
In particular, we know that massless vector quasinormal modes \cite{3}
are qualitatively similar to those of scalar or gravitational fields
\cite{3b}, except for limit of asymptotically high overtones: $Re
\omega$ approaches zero for vector field and is   $ln 3/8 \pi M $
for scalar and gravitational fields \cite{3c}.   

On the other hand, the massive term corrects the effective potential,
and for simplest case of scalar field it leads to the wave-equation
\begin{equation}
\Psi_{, r_{*} r_{*}} - \Psi_{,t t}- 
\left(1-\frac{2 M}{r}\right)\left(\frac{\ell (\ell+1)}{r^2} + \frac{2  M}{r^3} + m^2 \right) \Psi=0.   
\end{equation}  
Here one can take $m=0$ and re-cover the massless case. 
The corresponding quasinormal frequencies were found in \cite{3} for
massless and in \cite{4} for massive case. Massive scalar quasinormal modes
proved to show quite peculiar properties. Thus when one increases the
mass of the field $m$, the damping rates of the QN modes decrease 
strongly, so that existence of infinitely long living modes called
``quasi-resonances'' \cite{5} becomes possible. When increasing $m$, lower
overtones, one by one, transform into  ``quasi-resonances'', while all
the other higher modes remain ``ordinary'', i.e. damped \cite{5a}.
(For this to happen one needs to deal with relatively  large values of
$m$, so that in a more realistic picture considering backreaction of the 
scalar particle onto a black hole, existence of such quasi-resonances
is questionable.).  On the other hand, in 1992 Coleman, Preskill and Wilczek stated that
the classical vector monopole field is determined by the mass of the
field itself, and not by the mass of the black hole \cite{6}. This
stimulated the consideration of the late-time behaviour of the monopole massive
vector field in the Schwarzschild background in \cite{7} where the
suggestion of \cite{6} was supported by asymptotic treatment.

One of the earliest papers dealing with massive vector field
perturbations was that by Galtsov, Pomerantseva and Chizhov
\cite{Galtsov}, where it was shown that massive particles around black
holes have quasi-stationary states with hydrogen-like spectrum. That was
different from behaviour of the Proca field in Coulomb potential, where 
bounded states cannot be formed \cite{Tamm}. In the paper \cite{Galtsov}, the
perturbations equations were deduced for the first time, yet, as 
the system of equations for general value of multipole number $\ell$
cannot be decoupled, the solution was obtained in the region far from
a black hole \cite{Galtsov}. On the contrary in \cite{7}, the
perturbation equations were reduced to a single wave-like equation,
but only for the case of spherically symmetrical perturbations and
zero cosmological constant.

We are interested now to know what will happen with massive vector
perturbations in a black hole
background. In this case the situation is qualitatively different
from the known massless vector or massive scalar cases. First, the wave equation
for monopole massive vector perturbations cannot be reduced to that
one for the massless
vector field, just because the massless vector field does not have
radiative monopole. Another distinctive feature: the corresponding
effective potential is not positive definite everywhere outside the
black hole, so one must check the stability of perturbations.

The most unexpected feature we have found in the present paper is that 
when increasing the mass of the field, the lowest frequency and the higher overtones
behaviour is qualitatively different: the fundamental mode
decreases its damping rate what results in appearance of infinitely
long living modes, while, on the contrary, all higher modes decrease
their oscillation frequencies, leading to appearance of almost
non-oscillatory damping modes.

The paper is organised as follows: in Sec I we deduce the wave
equation for perturbations of the Proca field in the background of 
Schwarzschild,  Schwarzschild-de Sitter and  Schwarzschild-anti-de
Sitter black holes. In Sec. II the stability of monopole perturbations
is proved. Sec. III deals with quasinormal spectrum for massive vector
perturbations of Schwarzschild, and Schwarzschild-anti-de Sitter
backgrounds, including obtaining of the 
asymptotically high overtone limit. 
In the Conclusion we give a summary of obtained results.

\section{Perturbations of Proca field in a black hole background}

We shall consider here the  Schwarzschild black hole solution with a
$\Lambda$ - term, i.e. Schwarzschild,  Schwarzschild-de Sitter and
Schwarzschild-anti- de Sitter backgrounds in which the massive
vector field propagates. The black hole metric is given by
\begin{equation}
d s^2 = -f(r) d t^2 + f(r)^{-1} d r^2 + r^2 (d \theta^2 + sin^2 \theta d \phi^2),
\end{equation}
where
$$f(r)=\left(1-\frac{2 M}{r} -\frac{\Lambda r^2} {3} \right).$$

The vector field is described by a four-potential $A_{\mu}$, which is
supposed to interact with gravitational field minimally, i.e. the
field equations are generally-covariant analogs of the vector
field equations in Minkowskian space-time. Therefore, the Proca
equations
\begin{equation}
F^{\mu \nu}_{; \nu}- m^2 A^{\mu}=0, \quad F^{\mu \nu} = A_{\nu, \mu} - A_{\mu, \nu},
\end{equation}   
in curved space-time, read
\begin{equation}   
\frac{1} {\sqrt{-g}} ((A_{\sigma, \rho}- A_{\rho, \sigma}) g^{\rho
  \mu} g^{\sigma \nu} \sqrt{-g})_{,\nu} - m^2 A^{\mu } =0.
\end{equation}   
  
From here and on the coordinates $t$, $r$, $\theta$, and $\phi$ 
will be designated as $0$,$1$, $2$, and $3$ respectively. 

With respect to angular coordinates we imply adequate expansion into
spherical harmonics. Then, the field perturbations can be described by
four scalar functions of the radial coordinate and time $f^{\ell m}
(r, t)$, $h^{\ell m}(r, t)$, $k^{\ell m}(r,
t)$, and $a^{\ell m}(r, t)$:
\begin{equation}
A_{0} = f^{\ell m} (r, t) Y_{\ell m} (\theta, \phi),
\end{equation}   
\begin{equation}
A_{1} = h^{\ell m} (r, t) Y_{\ell m} (\theta, \phi),
\end{equation}   
\begin{equation}
A_{2} = k^{\ell m} (r, t) Y_{lm, \theta} (\theta, \phi) +  \frac{
  a^{\ell m}(r, t) Y_{\ell m}(\theta, \phi)} {sin
  \theta},
\end{equation}   
\begin{equation}
A_{3} = k^{\ell m} (r, t) Y_{\ell m, \phi} (\theta, \phi) - a^{\ell m}
(r, t) sin \theta Y_{\ell m, \theta}.   
\end{equation}  
 
Considering eq. (5) with $\mu=0$, $1$ and substituting Eqs. (6)-(9)
we arrive at the following equations:

\begin{equation}
\lambda ( k^{\ell m}_{, t} - f^{\ell m}) - ((h_{, t}^{\ell m}-
f^{\ell m}_{, r}) r^2)_{, r} f(r) + m^{2} r^{2} f^{\ell m} =0,
\end{equation} 
\begin{equation}
\lambda ( k^{\ell m}_{, r} - h^{\ell m}) - ((f_{, r}^{\ell m}-
h^{\ell m}_{, t}) r^2)_{, t} f(r)^{-1} + m^{2} r^{2} h^{\ell m} =0,
\end{equation} 
where $\lambda = \ell (\ell+1)$. Here we got rid of the function $a^{lm} (r,
t)$, so that the final perturbation dynamic can be described by the 
three independent functions of $r$ and $t$.   

The other two equations of (5), corresponding to  $\mu=2$, $3$, result in a
pair of equations with both even and odd spherical harmonics. 
Let us differentiate (10) with respect to  $r$
and (11) with respect to $t$.   Then, 
consider the particular case of spherically symmetrical
perturbations. Thereby taking $l=0$, i.e., implicitly, discarding all terms containing
derivatives with respect to angular variables, and introducing the new function 
\begin{equation}     
B = A_{r, t} - A_{t, r},
\end{equation} 
we obtain the following equation:
$$f(r) B_{, rr} - \frac{B_{, tt}}{f(r)} + \left(\frac{2} {r} - \frac{2 M}{r^2}
-  \frac{4 \Lambda r}{3} \right) B_{,
r} +$$ 
\begin{equation}  
  \left(\frac{8 M}{r^3} - \frac{2 \Lambda} {3} - \frac{2}{r^2} + m^2 \right) B = 0.
\end{equation} 

Assuming $B \sim e^{i \omega t}$, after introducing of $\Psi = B r$ and
using of the tortoise coordinate
$r_{*}$:  $d r_{*}= dr/f(r)$, we get the radial wave-like equation:

\begin{equation}
\frac{\partial^2 \Psi(r)}{\partial r_{*}^{2}} + \omega^2 \Psi - V(r)
\Psi=0,   
\end{equation} 
with the effective potential  
\begin{equation}
V(r) = \left(1-\frac{2 M}{r} -\frac{\Lambda r^2}{3}
\right) \left(\frac{2}{r^2} - \frac{6 M} {r^3} + m^2 \right).
\end{equation} 
When the $\Lambda$-term vanishes,  the wave equation (14, 15) reduces to
that obtained recently in \cite{7} with the help of 
the Newman - Penrose tetrad formalism. $\Lambda > 0 (\Lambda < 0)$
corresponds to asymptotically de-Sitter (anti-de Sitter) solutions.

Yet, for perturbations of general multi-polarity, all four equations
of (5) can be reduced to the matrix equation
for three scalar functions $\Psi_{\alpha}(r)$, $\alpha =0, 1, 2$,  
\begin{equation}
\frac{\partial^2 \Psi_{\alpha}(r)}{\partial r_{*}^{2}} +  M_{\alpha
\beta} (r, \omega) \Psi_{\beta}=0,   
\end{equation} 
and the matrix   $M_{\alpha \beta} (r, \omega)$ cannot be
diagonalized by the $r$-independent transformations of  the vector
$\Psi_{\alpha}(r)$, i.e. the set of equations (16) cannot be reduced 
to the wave-like equations (14).

\section{Effective potential and stability}

Even though we are limited now by spherically symmetric perturbations,
one can hope it is possible to judge about stability of the system
against massive vector field perturbations, because usually, 
if a system is stable against monopole perturbations, it is stable also against higher multipole
perturbations.  
The effective potential for different values of field
mass $m$ is given on Fig. 1., 2 and 3 for Schwarzschild,
Schwarzschild-de Sitter and Schwarzschild- anti-de Sitter black holes respectively.
To prove the stability of perturbations governed by the
wave equation (14, 15) we need to show, that the corresponding differential
operator
\begin{equation}
A= -\frac{\partial^2}{\partial r_{*}^{2}} + V(r)   
\end{equation} 
is positive self-adjoint operator in the Hilbert space of square
integrable functions of $r_{*}$, so that there is no normalizable growing
solution. This provides that all found quasinormal modes are damped. 
For massless scalar, vector and gravitational
perturbations (as well as for a massive scalar
perturbations) of a four-dimensional Schwarzschild, Schwarzschild-de
Sitter, and Schwarzschild - anti - de Sitter black holes, 
the effective potential is manifestly positive, and therefore the
positivity of the self-adjoint operator is evident.
As a result, the corresponding quasinormal modes for these cases 
are damped.  
Yet, for massive vector perturbations, as we see from  Fig. 1., 2 and
3, the effective potential has negative values near the event
horizon. Nevertheless, the effective potential is bounded from below and we can
apply here the method used in \cite{8}, which 
consists in extension of $A$ to a semi-bounded self-adjoint operator
in such a way, that the lower bound of the spectrum of the extension 
does not change. For this to perform, let us, following \cite{8}, introduce
the operator
\begin{equation}
D= \frac{\partial^2}{\partial r_{*}^{2}} + S(r),    
\end{equation} 
and we know that \cite{8}:
\begin{equation}
(\Psi, A \Psi)_{L^2} = - (\Psi^{*} D \Psi)_{boundary} + \int d r_{*}
(|D \Psi|^{2} + W |\Psi|^{2}), 
\end{equation}  
where
\begin{equation}
W=V + \left(1-\frac{2 M}{r} -\frac{\Lambda r^2}{3} \right) S'(r) - S^{2}(r).
\end{equation}     
Thus, we need to find the function $S(r)$ which would make the
effective potential $W$ positive. After investigation of
the form of the effective potential one can see that there is a set
of functions $S(r)$ satisfying this requirement. For instance, 
the function 
\begin{equation}
S(r)= \frac{1}{r} \left(1- \frac{2 M}{r} - \frac{\Lambda r^{2}}{3} \right)
\end{equation} 
creates the following potential:
\begin{equation}
W= \frac{4 (-3 M + r) \Lambda - m^{2} (6 M - 3 r - r^3 \Lambda)}{3 r}.
\end{equation} 
Using Mathematica, one can show that this potential is positive outside
the event horizon of a black hole. Thus a symmetric operator $A$ is
positive definite outside the black hole for positive and zero
cosmological constant, and so is the self-adjoint extension. Yet, for
the case of asymptotically anti-de Sitter black hole, the range of the 
tortoise coordinate is incomplete. At the same time, since the
effective potential is divergent at spatial infinity the Dirichlet
boundary conditions $\Psi(r=\infty) =0$ is physically motivated. 
Then, the boundary term in (19) does not contribute to the spectrum, and 
we obtain the positive self-adjoint extension of $A$. 
Thereby, we have proved that the  Schwarzschild, Schwarzschild-de Sitter and
Schwarzschild-anti- de Sitter space-times are stable against
monopole massive vector field perturbations. It means that there are no growing 
quasinormal modes in the spectra of these perturbations. In the next
section we shall compute the quasinormal modes for
the asymptotically flat and AdS cases, and  show, that all found modes
are damped implying the stability.

\begin{figure}
\resizebox{1\linewidth}{!}{\includegraphics*{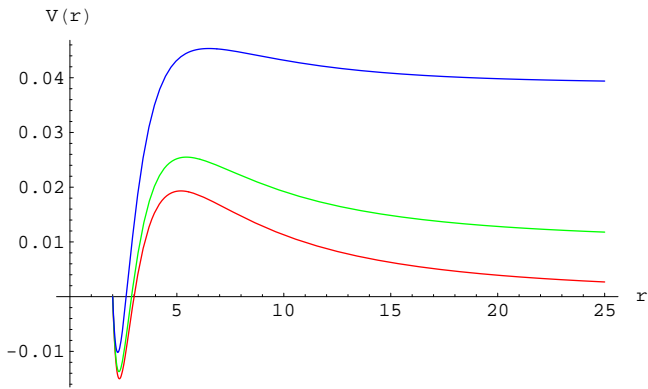}}
\caption{Effective potential as a function of radial coordinate for
  Schwartzschild black hole for different values of mass of the field:
  $m=0.01$ (bottom), $m=0.1$, and $m=0.2$ (top); $M=1$.} 
\label{near_extreme}
\end{figure}

\begin{figure}
\resizebox{1\linewidth}{!}{\includegraphics*{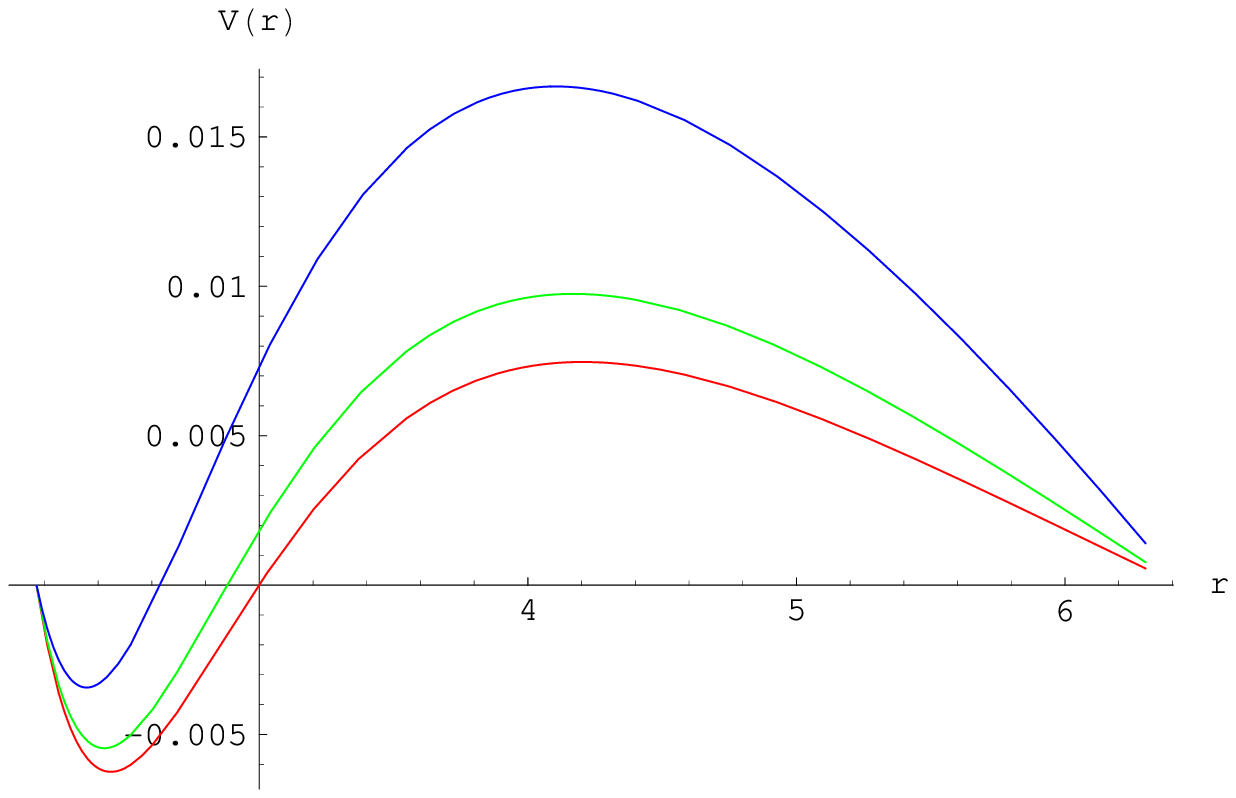}}
\caption{Effective potential as a function of radial coordinate for
  Schwartzschild-de Sitter black hole for different values of mass of the field:
  $m=0.01$ (bottom), $m=0.1$, and $m=0.2$ (top);
  $M=1$, $\Lambda=0.05$.} 
\label{near_extreme}
\end{figure}

\begin{figure}
\resizebox{1\linewidth}{!}{\includegraphics*{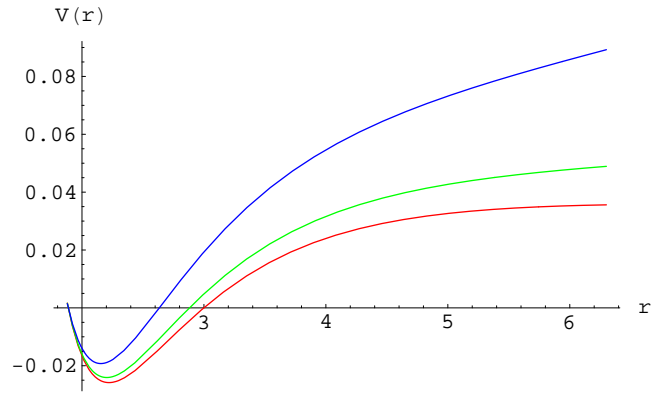}}
\caption{Effective potential as a function of radial coordinate for
  Schwartzschild- anti-de Sitter black hole for different values of mass of the field:
  $m=0.01$ (bottom), $m=0.1$, and $m=0.2$ (top); $M=1$, $\Lambda=0.05$.} 
\label{near_extreme}
\end{figure}

\section{Quasinormal modes}

We shall be restricted here  by consideration of quasinormal modes of asymptotically
flat and AdS black holes as those which are most physically
motivated. QNMs of asymptotically flat black holes may be observed by
future generation of gravitational antennas \cite{reviewQNM}, while  asymptotically AdS
black holes have direct interpretation in the
conformal field theory in the regime of strong coupling \cite{horowitz-hubeny}.

Let us start with asymptotically flat black holes.
The effective potential (17) approaches constant values both at event horizon and spatial
infinity in this case. Therefore the standard QN boundary conditions $\Psi \sim
e^{\pm i k_{\pm} r_{*}}, \quad r_{*} \pm \infty$ ($k_{+}=\omega, \quad
k_{-} =\sqrt{\omega^2 - m^2}$) are reasonable. More accurately, taking into
consideration the sub-dominant asymptotic term at infinity, the QN boundary conditions are
\begin{equation}
\Psi(r^*)\sim C_+
e^{i \chi r^*}r^{(i M m^2/\chi)}, \quad
(r,r^*\rightarrow+\infty),
\end{equation}
$$\chi = \sqrt{\omega^2-m^2}.$$
Note that the sign of $\chi$ is to be chosen to remain in the same
complex  plane quadrant as $\omega$.

Following the Leaver method, one can eliminate the singular factor
from $\Psi$, satisfying the in-going wave
boundary condition at the event horizon  and
(23) at infinity, and expand the remaining part into the Frobenius series that are
convergent in the region between the event horizon and the
infinity (see \cite{5a} for more details). The Frobenius series are:
$$\Psi(r) = e^{i \chi r}r^{(2i  M\chi+i  M
m^2/ \chi)}\left(1-\frac{2M}{r}\right)^{-2 i
M\omega } \times$$
\begin{equation}
\sum_na_n\left(1-\frac{2M}{r}\right)^n,
\end{equation}

Substituting (24) into (14) we find a three-term
recurrence  relation:
\begin{equation}
\alpha_0 a_1+ \beta_0 a_0 = 0, \quad  \alpha_n a_{n+1}+ \beta_n
a_n+\gamma_n a_{n-1}=0, \quad  n>0,
\end{equation}

Then, using algebra of continued fractions we can find the quasinormal
modes as those values of $\omega$ for which
$$\beta_n-\frac{\alpha_{n-1}\gamma_{n}}{\beta_{n-1}
-\frac{\alpha_{n-2}\gamma_{n-1}}{\beta_{n-2}-\alpha_{n-3}\gamma_{n-2}/\ldots}}=$$
\begin{equation}
\frac{\alpha_n\gamma_{n+1}}{\beta_{n+1}-\frac{\alpha_{n+1}\gamma_{n+2}}{\beta_{n+2}-\alpha_{n+2}\gamma_{n+3}/\ldots}}.
\end{equation}
As this procedure is described in many papers (see for instance  \cite{9} and references therein),
we shall not describe it here in details, and, go over directly to the obtained results.
We shall write $\omega = Re \omega + i Im \omega$, and the damped
modes should have $Im \omega < 0$

\begin{table}
\caption{First ten quasinormal modes for Schwartzschild black
 hole: $M=1$, $m=0.01$ and $m=0.1$ and $m=0.25$.} 
\label{sp1_figura1} 
\begin{ruledtabular}
\begin{tabular}{ccccccc}
\multicolumn{1}{c}{} &
\multicolumn{2}{c}{$ m=0.01$} &
\multicolumn{2}{c}{$ m=0.1$} &
\multicolumn{2}{c}{$ m=0.25$}\\ 
n  & Re($\omega_0$)  &  -Im($\omega_0$) & Re($\omega_0$) & -Im($\omega_0$)  &
Re($\omega_0$) & -Im($\omega_0$)
\\
\hline
\\
0 & 0.110523 & 0.104649  & 0.121577 & 0.079112 & 0.222081  & 0.012994 \\
1 & 0.086079 &  0.348013 &  0.082277 &  0.344140 & 0.062605 &  0.325191  \\ 
2 & 0.075725 &  0.601066 &  0.074036 & 0.599791 & 0.065511 &  0.592979 \\ 
3 & 0.070401 &  0.853671 &  0.069451 &  0.853002  & 0.064570 & 0.849359 \\
4 & 0.067068  & 1.105630 & 0.066451 & 1.105200  & 0.063243 & 1.102860 \\
5 & 0.064737 & 1.357140  &  0.064299 & 1.356830 & 0.062006 & 1.355170 \\
6 & 0.062991 &  1.608340  &  0.062660 & 1.608110 & 0.060925 & 1.606850\\
7 & 0.061619 &  1.859320 &  0.061359 &  1.859140  & 0.059991 & 1.858140 \\ 
8 & 0.060504  & 2.110150  & 0.060293 &  2.110001  & 0.059182  & 2.109180 \\ \
9 & 0.059575 &  2.360860  & 0.059400  & 2.360730  & 0.058475 & 2.360050\\

\end{tabular} 
\end{ruledtabular}
\end{table}

\begin{figure}
\resizebox{1\linewidth}{!}{\includegraphics*{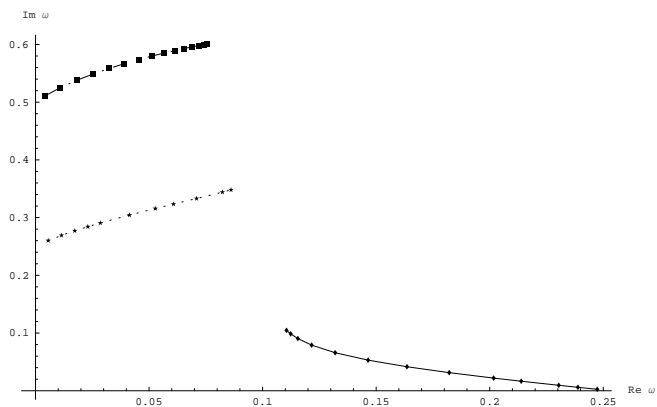}}
\caption{Imaginary part of $\omega$ as a function of real part of
  $\omega$ for first three overtones for increasing mass: $m \in (0.01,
  0.28)$ for $n=0$ (diamond), $m \in (0.01,
  0.48)$ for $n=1$ (star), $m  \in (0.01,
  0.75)$ for $n=2$ (box)} 
\label{near_extreme}
\end{figure}

First of all, let us look at the table I where first ten QNMs are
presented for three small (with respect to the black hole mass $M$) values 
of field mass $m$. Note, that as we 
consider a vector field minimally interacting with gravity, i.e.
back reaction of the vector field on the metric is not considered we 
cannot consider large values of $m$/$M$. We see from table I that
as the overtone number is increasing the difference between QNMs for 
different values of $m$ is decreasing and becomes small even at around tenth
overtone. Thus, one can conclude (and we check this by computing
high overtones), that high overtone behaviour does not depend on the
mass term $m$ coming into the effective potential (15), 
that is in agreement with previous study of high overtones for 
massive scalar field in \cite{5a}. Let us remind, that  
asymptotic limit of the QN spectrum for massive vector field does not
reduce to  that  for the massless case, because the effective
potential (15),  does not have physical meaning in the limit $m=0$. 
The most unexpected feature of the
quasinormal spectrum we found (see Fig. 4) is that the fundamental 
mode shows correlation with mass of the field
$m$, totally different from all the remaining higher overtones. Thus, 
as the mass $m$ is increasing, the real part of the fundamental
mode is increasing, while the imaginary part is falling off to tiny 
values, leading thereby to existence of the so-called quasi-resonant
modes, i.e. of infinitely long living oscillating modes \cite{5}, \cite{5a}.
On contrary, the second, third (see Fig. 4) and higher overtones
have their real part decreasing to tiny values, and, the imaginary part
is growing when the mass $m$ is growing. Thus higher overtones can
lead to existence of almost pure imaginary modes which just damp
without oscillations. 
We do not know examples of such a different
behaviour between the fundamental mode and higher overtones, for massless fields
of any spin \cite{10} or for massive scalar field \cite{4},
\cite{5}, \cite{5a}, at least for
asymptotically flat or de Sitter black holes. 
The infinitely long living modes can exist for massive scalar field
perturbations \cite{5}, but for {\it all} modes \cite{5a}, not only for
the fundamental one. Note however, that
for massless vector perturbations of asymptotically AdS black holes under
Dirichlet boundary conditions, the fundamental mode is pure
imaginary (see for instance \cite{11} and references therein), 
what represents the hydrodynamic mode in the dual conformal field
theory \cite{12}. So, the qualitative difference between fundamental and 
higher overtones is not absolutely new phenomena, yet, completely
unexpected for asymptotically flat space-times.
Note also, that despite the fact that we have two tendencies: approaching  $Re \omega$
the constant value $\log 3 /8 \pi M$ , when $n$ is growing,  and at the same time
approaching zero, when $m$ is growing, there is no
contradiction:  to approach the limit  $\log 3 /8 \pi M$, real part of $\omega$
should increase again after some certain $n = n_{c}$ \cite{referee}. We can observe
it on the Table I. (third column), where we can observe the ``local
maximum'' of $Re \omega$ at $n=2$.

The modes in Table I and Fig. 4 were found with the help of the above
described Frobenius technique. For lower overtones, one can use,
alternatively, the WKB approach suggested in \cite{13} and
consequently developed to  3th \cite{13a} and 6th  \cite{13b} WKB
orders beyond the eikonal approximation. The WKB formula has been used
recently in a lot of papers \cite{WKBuse} and comparison with accurate 
numerical data in many cases \cite{WKBcompare} shows good accuracy of the WKB 
formula up to the 6th WKB order.    
Here we can compare the results with WKB values, but only for the fundamental overtone, because for higher ones:
$n > \ell=0$, and the WKB method cannot be applied.

Note also, that  an effective potential takes negative values near the event
horizon, and, the WKB formula does not take into consideration
 ``sub'' scattering by the local 
minimum of the potential and should not be so accurate
as in the case of the ordinary positive definite potential.
For example, for $m=0.01$ we get  $0.110523 - 0.104649 i$ with the help of the  
Frobenius method, and $0.1195 - 0.0871 i$ by WKB formula \cite{13}. The
larger $m$ is, the worse convergence of the WKB method. Generally we see
that the accuracy of WKB approach is not satisfactory here because
the WKB formula is actually good only for $\ell > n$.

Now let us  go over to asymptotically high overtones. It is known
\cite{5a},  
that the mass term does not change the infinitely high overtone asymptotic
of the Schwarztschild black hole. Thus it is natural to expect
that the same will take place for a  massive vector field. Yet,
as there are no  monopole dynamical degrees of freedom for massless
vector perturbations, we cannot formally take $m=0$ in the
considered effective potential. Therefore, using the Nollert's method \cite{14}, 
we computed numerically high overtones for non-vanishing values of $m$
(see Fig. 5). From Fig. 5 one can learn that as $n$ is growing, the 
real part approaches $ln 3/8 \pi M $ , while the spacing in imaginary part
approaches constant:
\begin{equation}
Re \omega_{n} \rightarrow \frac{ln 3}{8 \pi M}, \quad Im \omega_{n}
\rightarrow \frac{(2 n-1)}{8 M}, \quad n \rightarrow \infty.
\end{equation}

\begin{figure}
\resizebox{1\linewidth}{!}{\includegraphics*{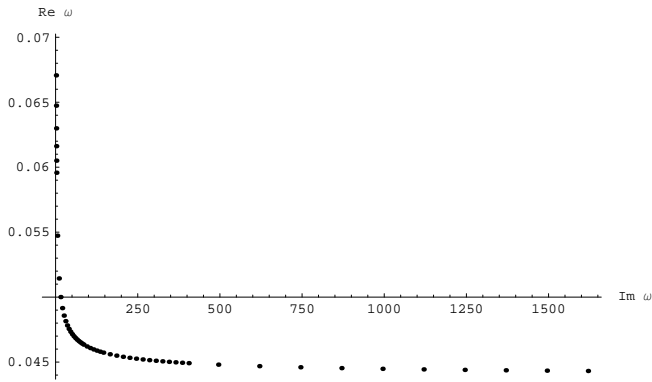}}
\caption{Imaginary part of $\omega$ as a function of real part of
  $\omega$ for high overtone behaviour: $m=0.01$, $M=1$.} 
\label{1}
\end{figure}

It is different from the asymptotic limit which takes place for higher multipole
perturbations of massless vector field \cite{15}: because for the latter
case the real part of $\omega$ asymptotically approaches zero \cite{15}.
To see better that in the obtained plot $Re \omega$ approaches  $ln 3/8
\pi M $ let us make fit on values $n=1000, 1500, 2000,....6000$.
In a similar fashion with Nollert's approach, we see that fit in powers
of $1/\sqrt{n}$ is better then in powers of  $1/n$, and gives
\begin{equation}
Re \omega_{n} \approx 0.04372 + \frac{0.04780}{\sqrt{n}}+
\frac{0.03353}{n}, \quad n \rightarrow \infty.
\end{equation}  
This is very close to  $ln 3/(8 \pi M) \approx 0.04373$. 
When increasing the number of overtones, the obtained fit is closer to  $ln 3/(8 \pi M)$.

Following the arguments of \cite{15}, it is straightforward to reproduce numerically
obtained asymptotic (28) in an analytical way.
For this it is enough to remember that the effective potential (15), has
the following asymptotic behaviour in the origin :
\begin{equation}
V(r) \rightarrow \frac{12 M^2}{r^4}, \quad r \rightarrow 0, 
\end{equation}
and at the event horizon
\begin{equation}
V(r) \rightarrow const (r- 2 M) + O((r- 2 M)^{2}), \quad r \rightarrow 2 M. 
\end{equation}
Therefore the general asymptotic solution near the origin is
\begin{equation}
\Psi(r_{*}) = c_{1} \sqrt{\omega r_{*}} J_{1}(\omega r_{*}) +  c_{2} \sqrt{\omega r_{*}} J_{-1}(\omega r_{*}), \quad r \rightarrow 0.    
\end{equation}
Then, repeating all relevant steps of \cite{15} and taking into
account that near  the event horizon the wave function has the
following asymptotic  
\begin{equation}
\Psi(r_{*}) \sim e^{2 M i  \omega  ln (r_{*}(r)- r_{*}(r=2 M))}, \quad r \rightarrow 2 M,
\end{equation}          
and equating the two monodromy (which look similar to those in
\cite{15}) one gets (28).

We see that the high overtone asymptotic (28) is the same as for
gravitational perturbations. This is easily understood, because 
the effective potential looks like that for gravitational perturbations
with formally taken $\ell=1$ plus massive term times f(r). Then, as we
have shown here for vector and in \cite{5a} for scalar fields, the
massive term does not contribute in high overtone asymptotic.



The quasinormal behaviour of SAdS black holes is essentially dependent
on radius of a black hole: one can distinguish the three regimes of 
large ($r_{+} >> R$), intermediate  ($r_{+} \sim R$), and small
($r_{+} << R$) AdS black holes. From detailed previous study of
massless fields, one
can learn that QNMs of large AdS black holes are proportional 
to the black hole radius, and therefore to the temperature \cite{horowitz-hubeny}.
QNMs of intermediate AdS black holes do not show simple linear
dependence on radius \cite{horowitz-hubeny} . Finally, QNMs of small AdS black holes
approach normal modes of empty AdS space-time \cite{last}. In the limit of
asymptotically high damping, QNMs show equidistant spectrum 
with the same spacing between nearby modes for different massless
fields (scalar, electromagnetic and gravitational) \cite{11}.

Using the Horowitz-Hubeny method, we obtain the
quasinormal frequencies for SAdS black hole  numerically. As this method is
described in a lot of recent works, we shall outline only the key
points of it here. The Schwarzschild-AdS metric function can be written in the
form
\begin{equation}
f(r) = 1 - \frac{r_{0}} {r} + \frac{r^2}{R^2},
\end{equation}
where  $R$ is the anti-de Sitter radius. 
The corresponding effective potential is divergent at infinity and is
polynomial function of $r$. Therefore, one can expand the wave
function  $\Psi$ near the event horizon in the form:
\begin{equation}
\Psi(x) = \sum_{n=0}^{\infty} a_{n} (x-x_{+})^{n}, \quad x_{+}=1/r_{+}. 
\end{equation}

Here $r_{+}$ is the largest of the zeros of the metric function
$f(r)$. The Dirichlet boundary conditions, we shall use here, imply that
\begin{equation}
|\Psi(r=\infty)| =0. 
\end{equation}
Then we need to truncate the sum (34) at some large $n=N$, 
in order to observe the convergence of the values of the root of the
equation (35) $\omega$  to some true quasinormal frequency.

The fundamental quasinormal frequencies are shown in Table II for 
large, intermediate, and small SAdS black holes for different values of $m$.  
From Table II one can see that both real and imaginary parts of the
quasinormal frequency are increasing when  the mass of
the field is growing.


\begin{widetext}

\begin{table}
\caption{Fundamental quasinormal modes for large ($r_{+}=100 R$)
    intermediate ($r_{+}=1 R$), and small ($ r_{+}=1/10 R $)
    Schwartzschild-anti-de Sitter black hole.} 
\label{sp1_figura1} 
\begin{ruledtabular}
\begin{tabular}{ccccccc}
\multicolumn{1}{c}{} &
\multicolumn{2}{c}{$r_{+}=100 R$} &
\multicolumn{2}{c}{$r_{+}=1 R$} &
\multicolumn{2}{c}{$r_{+}=1/10 R$}\\ 
m  & Re($\omega_0$)  &  -Im($\omega_0$) & Re($\omega_0$) & -Im($\omega_0$)  &
Re($\omega_0$) & -Im($\omega_0$)
\\
\hline
\\
0.01 &  184.959733 & 266.38559  & 2.798314 & 2.671325 & 2.6929  &  0.1010 \\
0.05 & 185.109096 &  266.671681 & 2.800496 &  2.674197 & 2.6949 & 0.1012 \\ 
0.1 & 185.571219 &  267.521739 & 2.807245 & 2.683084 & 2.700 & 0.103 \\ 
0.15 & 186.325972 & 268.912194 & 2.818276 &  2.697634 & 2.709 & 0.1035\\
0.2 & 187.352413 & 270.806932 & 2.833289  & 2.717471  & 2.7247 &  0.1039 \\
0.25 & 188.625045 & 273.161253 & 2.851928   & 2.742088  & 2.7416 & 0.1055 \\
\end{tabular} 
\end{ruledtabular}
\end{table}

\end{widetext}

Finally, let us find the normal modes of pure AdS space-time
for the case of massive vector field. The metric function $f(r)$ of pure AdS
space-time has the form:
\begin{equation}
f(r) = 1 + \frac{r^2} {R^2}.
\end{equation}
We can put the anti-de Sitter radius to be $R=1$ in further
calculations. The tortoise coordinate is connected with the Schwarzschild
radial coordinate by the relation: 
\begin{equation}
r = \tan r^{*}.
\end{equation} 
Then, the effective potential has the form:
\begin{equation}
V = \frac{2}{sin^{2} r^{*}} + \frac{m^2} {cos^{2} r^{*}}.
\end{equation} 
Let us introduce a new variable 
\begin{equation}
z = \sin^{2} r^{*}.
\end{equation}
Then the wave equation can be written in the form:
\begin{equation}
4 z (1-z) \Psi_{, z z} (z) + 2 (1- 2 z) \Psi_{, z} (z) +
\left(\omega^2 - \frac{2}{z} -\frac{m^2}{1-z} \right) \Psi = 0.
\end{equation}
After introducing a new function 
$$\Psi = \Phi z^{\alpha} (1-z)^{\beta}, $$
the wave equation takes the form:
\begin{widetext}
$$
z (1-z) \Phi_{, z z} (z) + \left(\frac{1}{2} + 2 \alpha - (2 \alpha +
2 \beta +1) z\right) \Phi_{, z} (z) + $$
\begin{equation}
+ \left(\frac{2 \alpha (\alpha - 1) + \alpha - 1}{2 z} -\frac{2 \beta
(\beta - 1)+ \beta  - (m^2/2)}{2 (1-z)} + \frac{\omega^2}{4}  -
(\alpha + \beta)^{2}  \right) \Phi = 0.
\end{equation}
\end{widetext}

We are in position now to choose the values of $\alpha$ and $\beta$, so that
the terms proportional to $1/z$ and $1/(1-z)$ vanish. The general
solution is  
\begin{widetext}

$$\Psi = C_{1} z^{(1/2) - \alpha} (1-z)^{\beta} _{2} F_{1} (\frac{1}{2} - \alpha
+ \beta - \frac{\omega}{2},\frac{1}{2} - \alpha
+ \beta + \frac{\omega}{2}, \frac{3}{2} - 2 \alpha,
z) +$$ 
\begin{equation}
+ C_{2} z^{\alpha} (1-z)^{\beta} _{2} F_{1} (\alpha + \beta -
\frac{\omega}{2},\alpha + \beta +\frac{\omega}{2}, \frac{1}{2} + 2 \alpha, z),
\end{equation}
\end{widetext}
where constants $C_{1}$, $C_{2}$ may be complex.  
We require regularity of the solution at the origin $z=0$, $C_{1} =
0$, and, the vanishing of the wave function at spatial infinity
implies
\begin{equation}
_{2} F_{1} (\alpha + \beta -
\frac{\omega}{2},\alpha + \beta +\frac{\omega}{2}, \frac{1}{2} + 2
\alpha, z) = 0.
\end{equation}

Note also, that the choice $\alpha =1$, $\beta = \frac{1}{4}(1 -
\frac{1}{4}\sqrt{1+ 4 m^2})$ corresponds to the above boundary
conditions both at infinity and at the event horizon.

Now, it is not hard to see that $\omega$ has the form:
\begin{equation}
\omega_{n} =\sqrt {|\Lambda|/3} \left(2 n + 3 + \frac{1}{4}(1 -
\frac{1}{4}\sqrt{1+ 4 m^2})\right).
\end{equation} 

This is different from the AdS normal modes for massless vector
field $\omega_{n} =\sqrt {\Lambda /3} (2 n + 2 + \ell)$ \cite{15},
where $\ell = 1, 2, ...$, i.e. monopole perturbations are not dynamical.

Note, that it is expected that similar to scalar field behaviour
\cite{last}, the massive vector quasinormal modes of SAdS black holes
should approach their pure AdS values (44) as the mass of the black
hole goes to zero. For the fundamental mode, we see that according to formula (44), 
for $m=0.1$, one has $\omega = 3.004$, and, according to the extrapolation of
the data in Fig. VI obtained numerically with the help of Horowitz - Hubeny
method,  $\omega$, indeed, approaches some constant value close to
 $3$.
Unfortunately we cannot check the accurate numerical correspondence to the formula
(44), because the series (34) converges very slowly for small black
hole radius, and therefore one needs enormous computer time to achieve
the regime of very small black holes.

\begin{figure}
\resizebox{1\linewidth}{!}{\includegraphics*{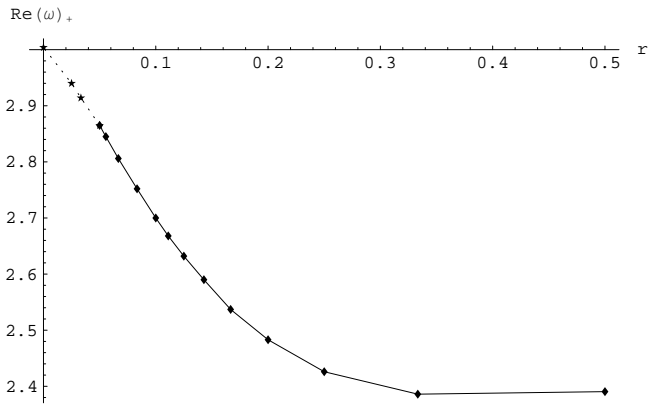}}
\caption{Real part of $\omega$ as a function of the black hole radius
$r_{+}$ for small AdS black holes: $m=0.1$, $R=1$.} 
\label{1}
\end{figure}

\begin{widetext}

\begin{table}
\caption{Higher overtones  for large ($r_{+}=100 R$)
    Schwartzschild-anti-de Sitter black hole.} 
\label{sp1_figura1} 
\begin{ruledtabular}
\begin{tabular}{ccccccc}
\multicolumn{1}{c}{} &
\multicolumn{2}{c}{$m=0.1 $} &
\multicolumn{2}{c}{$m=0.2 $} &
\multicolumn{2}{c}{$m=0.3 $}\\ 
n  & Re($\omega_0$)  &  -Im($\omega_0$) & Re($\omega_0$) & -Im($\omega_0$)  &
Re($\omega_0$) & -Im($\omega_0$)
\\
\hline
\\
1 & 316.780  &  492.755  & 318.601  &  495.990 & 321.437 & 501.024  \\
2 & 447.102 &  717.863 & 448.936  &  721.089  & 451.792  &  726.109 \\ 
3 & 577.201  & 942.921  & 579.043  & 946.141 & 581.913 & 951.155 \\ 
4 & 707.218  & 1167.952   & 709.064 &  1171.171 & 711.940 & 1176.185 \\
5 & 837.194  & 1392.973 & 839.044  & 1396.191 & 841.925 & 1401.203 \\
6 & 967.150   & 1617.987 & 969.002   & 1621.204  & 971.886 & 1626.215  \\
7 & 1097.093  & 1842.997 & 1098.946 &   1846.214 & 1101.832 & 1851.226 \\
8 & 1227.027  & 2068.006  & 1228.881  &  2071.222 & 1231.77  & 2022.23 \\
9 & 1356.956 & 2293.009 & 1358.811  &  2296.230  & 1361.70 & 2301.24 \\
\end{tabular} 
\end{ruledtabular}
\end{table}

\end{widetext}

Numerical data for high overtones, in the regime of large black holes,
is shown in Table III. 
There one can see that, indeed, in concordance with analytical formula (34), one has
\begin{equation}
\frac{\omega_{n+1}-\omega_{n}}{r_{+}}  \approx 1.29 - 2.25 i.
\end{equation}
At a sufficiently high $n$, the above formula is valid independently
of the value of the mass field $m$. It is also valid for any large
$r_{+}$ because in the regime of large black holes the QNMs are
proportional to the black hole radius $r_{+}$ for massive fields as well.  
One could say that a quasinormal mode at high damping consist of two
contributions. One is proportional to an overtone
number $n$ and thereby equals to a spacing between  nearby modes;
it is called ``gap''. Another contribution does not depend on $n$ in the limit
$n \rightarrow \infty$, called ``offset''. Thus, one has   
\begin{equation}
\omega_{n} = [offset] + [gap] n, \quad  [gap] = 2 \sqrt{3}
\pi R^{2} e^{- i \pi/3}/9,   
\end{equation}
where $R$ is the anti-de Sitter radius, i.e. at high overtones, the
spectrum is equidistant with spacing which does not depend
on $m$ , and is the same as for gravitational or massless vector
perturbations. This should be true also for intermediate and small AdS
balck holes, yet to check this numerically one needs considerable
computer time, because of the slow convergence of the series for small
black holes.

All numerical computations in this paper were made with the help of
{\it Mathematica}.

\section{Conclusion}
   
Fortunately the monopole perturbations of the Proca field in the Schwarzschild-(A)dS
black hole background can be reduced to the wave-like equation with some effective
potential. Even though the effective potential is not positive
definite everywhere outside black hole, we have proved that 
spherically symmetrical perturbations of massive vector field is stable, i.e. there are not
growing modes in this case. This is confirmed by numerical
computations of the QNMs spectrum, which is done for  Schwarzschild
and  Schwarzschild-AdS black holes.  Quite unexpected property, we found,
is that the behaviour of the fundamental mode and all higher
overtones (for asymptotically flat case) are qualitatively different: 
when increasing the field mass $m$, the damping rate of the
fundamental mode goes gradually to zero, leading to appearing of
infinitely long living mode, while all higher overtones, on contrary, 
decrease their $Re \omega$ what results in existence of almost pure
imaginary modes, i.e. damping modes without oscillations.   

Asymptotics of infinitely high overtones for Schwarzschild balck holes are the same as for
corresponding gravitational (massless) perturbations. In particular, for
Schwarzschild black hole, real oscillation frequency approaches $ln
3/8 \pi M $, while damping rates become equidistant with spacing equal $1/4 M$.       
This value of  high damping asymptotic, which coincides with that for
massless scalar and gravitational fields for  Schwarzschild black
holes can be easily explained by two factors: 1) the mass term does
not contribute to the limit of infinite damping of the quasinormal
spectrum, and 2) when formally taking the limit $m=0$ in the effective
potential which governs the evolution of massive vector perturbations,
one has the potential which looks qualitatively like that for gravitational
perturbations.

For asymptotically AdS black holes the quasinormal spectrum is equidistant at high overtones
with spacing which does not depend on the mass of the field.

\begin{acknowledgments}
I would like to acknowledge A. Zhidenko for useful discussions and 
providing me with a copy of Ref.13. This work was  supported 
by \emph{Funda\c{c}\~{a}o de Amparo \`{a} Pesquisa do Estado de S\~{a}o Paulo (FAPESP)}, Brazil.

\end{acknowledgments}

\end{document}